\documentclass[a4paper]{jpconf}
\usepackage{graphicx} % Required for inserting images
\usepackage{iopams}
\usepackage{amsmath}
\usepackage[hidelinks]{hyperref}
\usepackage[noabbrev]{cleveref}
\newcommand{\ba}{\begin{eqnarray}}
\newcommand{\ea}{\end{eqnarray}}

\begin{document}

\title{Heavy $\Xi_{c/b}$ and $\Xi'_{c/b}$ baryons in the quark model}

\author{E. Ortiz-Pacheco}
\address{Jozef Stefan Institute, Jamova 39, 1000 Ljubljana, Slovenia}
\ead{emmanuel.ortiz@ijs.si}

\author{R. Bijker}
\address{Instituto de Ciencias Nucleares, Universidad Nacional Aut\'onoma de M\'exico,\\
Coyocac\'an, 04510 Ciudad de M\'exico, M\'exico}
\ead{bijker@nucleares.unam.mx}

\begin{abstract}
We present a study of the low-lying $\Xi_{c/b}$ and $\Xi'_{c/b}$ baryon states based on masses, strong and radiative decays. In a harmonic oscillator quark model analysis it is found that the majority of the detected states can be interpreted as $S$- and $P$-wave excitations.
\end{abstract}

\section{Introduction}
\label{sec:introduction}

According to the Particle Data Group compilation \cite{PDG20}, there are 15 single-charm cascade baryons, namely $\Xi_{c/b}$ and $\Xi'_{c/b}$, whose angular momentum and parity for certain states have yet to be determined. The LHCb collaboration has recently observed three negative parity states of $\Xi^0_c$, specifically $\Xi^0_c(2923)$, $\Xi^0_c(2939)$, and $\Xi^0_c(2965)$, which follow a mass rule of roughly 126 MeV with respect to the excited negative parity $\Omega^0_c$ states. Therefore, it is suggested that these three new negative parity excited $\Xi^0_c$ states belong to the same flavor $SU_{\rm f}(3)$ multiplet as the negative parity excited $\Omega^0_c$ states, a flavor sextet $\bf 6$. Additionally, it is assumed that $\Xi_{c}(2965)$ is the same state as the previously observed $\Xi_{c}(2970)$.

This observation made by the LHCb collaboration is important as it sheds light on the fundamental binding mechanism of matter at a basic level and emphasizes the significance of comprehending the interactions between quarks. The quantum numbers of the states are assigned using the quark model systematics, and the equal spacing mass rule for the $SU_{\rm f}(3)$ ground states in the Gell-Mann-Okubo and G\"ursey-Radicati mass formulas is similar to the spacing of the three new negative parity excited $\Xi^0_c$ states. Nonetheless, further research is required to validate these findings and gain a more in-depth understanding of the quark structure of hadrons.

\section{$\Xi_{c/b}$ and $\Xi'_{c/b}$ baryons}

\begin{table}[t]
\centering
\caption{Classification of highest charge state of sextet $\Xi'_{c/b}$ baryons (top) and anti-triplet $\Xi_{c/b}$ baryons (bottom).}
\label{XiQ}
\vspace{10pt}
\begin{tabular}{ccccc}
\hline\hline
\noalign{\smallskip}
State & Wave function & $(n_{\rho},n_{\lambda})$ & $L^P$ & $J^P$ \\
\noalign{\smallskip}
\hline
\noalign{\smallskip}
$^{2}\Xi'_Q$ & $\frac{1}{\sqrt{2}}(us+su)Q 
\left[ \psi_{0} \chi_{\lambda} \right]$ 
& $(0,0)$ & $0^+$ & $\frac{1}{2}^+$ \\ 
\noalign{\smallskip}
$^{4}\Xi'_Q$ & $\frac{1}{\sqrt{2}}(us+su)Q 
\left[ \psi_{0} \chi_{S} \right]$ 
& $(0,0)$ & $0^+$ & $\frac{3}{2}^+$ \\ 
\noalign{\smallskip}
$^{2}\lambda(\Xi'_Q)_J$ & $\frac{1}{\sqrt{2}}(us+su)Q 
\left[ \psi_{\lambda} \chi_{\lambda} \right]_J$ 
& $(0,1)$ & $1^-$ & $\frac{1}{2}^-$, $\frac{3}{2}^-$ \\ 
\noalign{\smallskip}
$^{4}\lambda(\Xi'_Q)_J$ & $\frac{1}{\sqrt{2}}(us+su)Q 
\left[ \psi_{\lambda} \chi_{S} \right]_J$  
& $(0,1)$ & $1^-$ & $\frac{1}{2}^-$, $\frac{3}{2}^-$, $\frac{5}{2}^-$ \\ 
\noalign{\smallskip}
$^{2}\rho(\Xi'_Q)_J$ & $\frac{1}{\sqrt{2}}(us+su)Q 
\left[ \psi_{\rho} \chi_{\rho} \right]_J$  
& $(1,0)$ & $1^-$ & $\frac{1}{2}^-$, $\frac{3}{2}^-$ \\ 
\noalign{\smallskip}
\hline
\noalign{\smallskip}
$^{2}\Xi_Q$ & $\frac{1}{\sqrt{2}}(us-su)Q 
\left[ \psi_{0} \chi_{\rho} \right]$ 
& $(0,0)$ & $0^+$ & $\frac{1}{2}^+$ \\ 
\noalign{\smallskip}
$^{2}\lambda(\Xi_Q)_J$ & $\frac{1}{\sqrt{2}}(us-su)Q 
\left[ \psi_{\lambda} \chi_{\rho} \right]_J$ 
& $(0,1)$ & $1^-$ & $\frac{1}{2}^-$, $\frac{3}{2}^-$ \\ 
\noalign{\smallskip}
$^{2}\rho(\Xi_Q)_J$ & $\frac{1}{\sqrt{2}}(us-su)Q 
\left[ \psi_{\rho} \chi_{\lambda} \right]_J$  
& $(1,0)$ & $1^-$ & $\frac{1}{2}^-$, $\frac{3}{2}^-$ \\ 
\noalign{\smallskip}
$^{4}\rho(\Xi_Q)_J$ & $\frac{1}{\sqrt{2}}(us-su)Q 
\left[ \psi_{\rho} \chi_{S} \right]_J$  
& $(1,0)$ & $1^-$ & $\frac{1}{2}^-$, $\frac{3}{2}^-$, $\frac{5}{2}^-$ \\ 
\noalign{\smallskip}
\hline\hline
\end{tabular}
\end{table}

The $\Xi_{c/b}$ and $\Xi'_{c/b}$ baryons can be described in the quark model as $qqQ$ configurations, two light quarks and one heavy quark. The light quarks $q$ correspond to the $u$, $d$ or $s$ flavors, whereas the heavy quark $Q$ can be either $c$ or $b$. The masses of the singly-heavy cascade baryons are calculated in a harmonic oscillator quark model with spin, spin-orbit, isospin, and flavor interactions as follows
\ba
H = H_{\rm ho} + A \, \vec{S} \cdot \vec{S} + B \, \vec{L} \cdot \vec{S} 
+ E \, \vec{I} \cdot \vec{I} + G \, C_{2SU_{\rm f}(3)} ~,
\label{hosc}
\ea
where the Hamiltonian of the harmonic oscillator for this system \cite{Isgurandkarl(1978)}
\ba
H_{\rm ho} &=& \sum_i \left( m_i + \frac{p_i^2}{2m_i} \right) 
+ \frac{1}{2} C \sum_{i<j} (\vec{r}_i - \vec{r}_j)^2 
\nonumber\\ 
&=& M + \frac{P^2}{2M} + \frac{p_{\rho}^2}{2m_{\rho}}  + \frac{p_{\lambda}^2}{2m_{\lambda}}
+ \frac{1}{2} m_{\rho} \omega_{\rho}^2 \rho^2  
+ \frac{1}{2} m_{\lambda} \omega_{\lambda}^2 \lambda^2 ~,
\ea
is expressed in terms of the Jacobi coordinates 
\ba
\vec{\rho} &=& (\vec{r}_1 - \vec{r}_2)/\sqrt{2} ~,
\nonumber\\
\vec{\lambda} &=& (\vec{r}_1 + \vec{r}_2 - 2\vec{r}_3)/\sqrt{6} ~, 
\nonumber\\
\vec R   &=& \frac{m_q(\vec r_1 + \vec r_2) + m_Q\vec r_3}{2m_q+m_Q} ~,
\ea
and their conjugate momenta. The reduced masses are given by $m_{\rho}=(m_{u/d}+m_s)/2$, and $m_{\lambda}=3 m_{\rho}m_Q/M$, with $M=2m_{\rho}+m_Q$, and the oscillator frequencies by  $\omega_{\rho}=\sqrt{3C/m_{\rho}}$ and $\omega_{\lambda}=\sqrt{3C/m_{\lambda}}$.
The baryons $\Xi_{c/b}$ and $\Xi'_{c/b}$ are isospin doublets with $I=1/2$. Since both the isospin $I$ and the sum of the quark masses $M$ have the same value for all baryons, they are absorbed into $M_0$. The mass formula is given by
\ba
E &=& M_0( ^2\Xi_Q ) + \hbar \omega_{\rho} n_{\rho} 
+ \hbar \omega_{\lambda} n_{\lambda} + A \left[ S(S+1) - \frac{3}{4} \right]+ B \frac{1}{2} \left[ J(J+1) - L(L+1) - S(S+1) \right] 
\nonumber\\
&&  + G \frac{1}{3} \left[ p(p+3)+q(q+3)+pq-4 \right] ~,
\label{MassFormula}
\ea 
where $S$, $L$, and $J$ are the spin, orbital angular momentum and total angular momentum, respectively.  

We only consider baryons associated with the ground-state configuration $\psi_0$ with $(n_{\rho},n_{\lambda})=(0,0)$, and with one quantum of excitation, either in the $\lambda$ mode, $\psi_{\lambda}$, with $(n_{\rho},n_{\lambda})=(0,1)$, or in the $\rho$ mode, $\psi_{\rho}$, with $(n_{\rho},n_{\lambda})=(1,0)$. A summary of the baryon wave functions together with their quantum numbers is presented in Table \ref{XiQ}.  

\section{Masses and decay widths}
\label{phpl}

\begin{figure*}[h]
\centering
\begin{minipage}{\linewidth}
\includegraphics[width=0.47\linewidth]{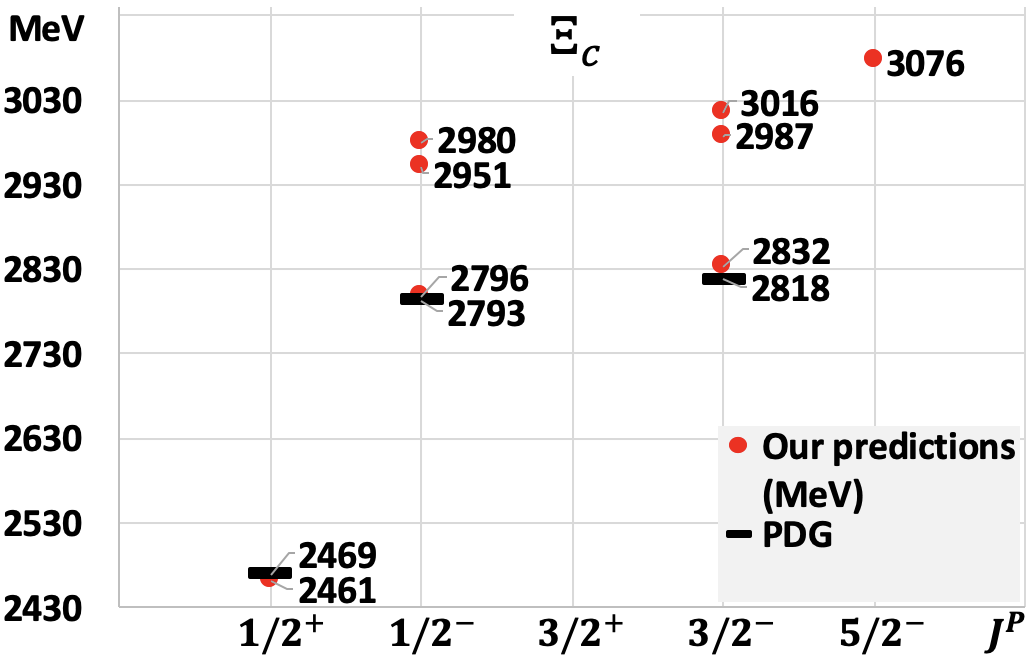}
\hfill
\includegraphics[width=0.47\linewidth]{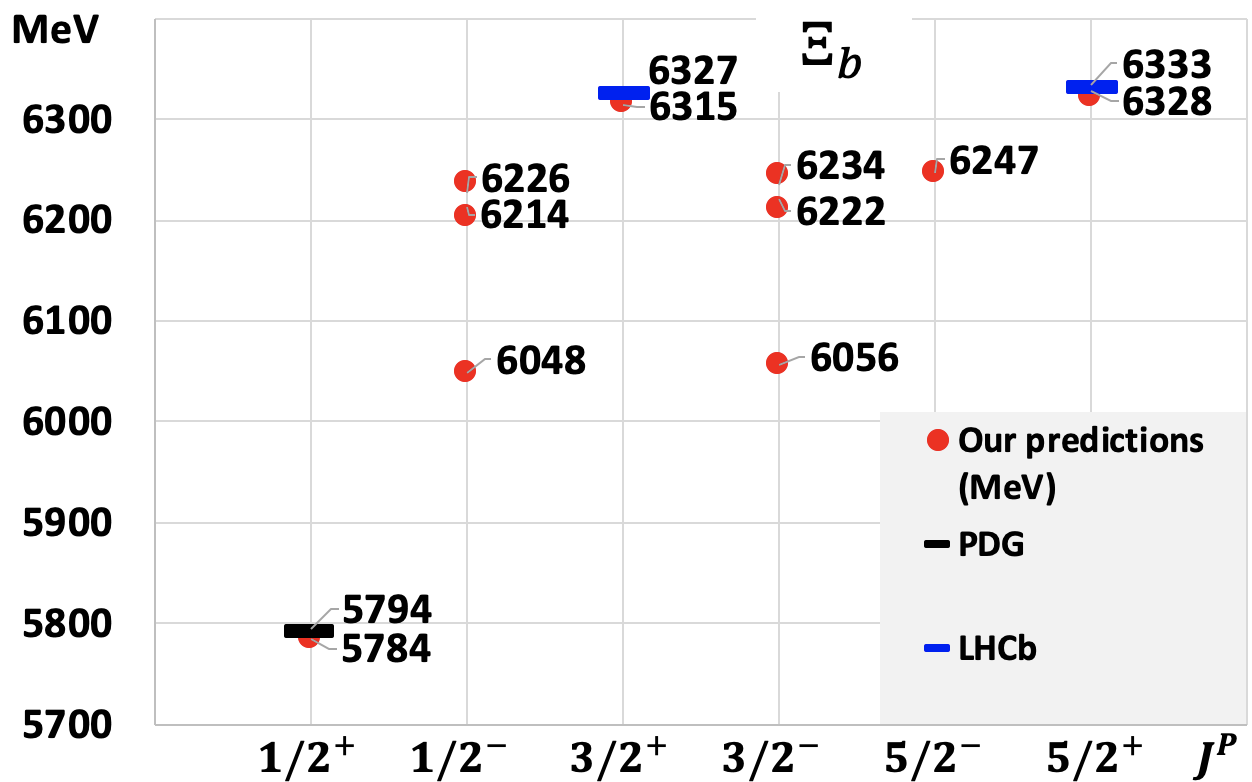}
\end{minipage}
\caption{The mass spectra of $\Xi_c$ (left) and $\Xi_b$ (right)  along with tentative quantum number assignments, are shown. The theoretical predictions (represented by red dots) are compared with experimental results obtained by the LHCb collaboration \cite{PhysRevLett.124.222001} \cite{PhysRevLett.128.162001} (depicted as blue lines), as well as with the Particle Data Group compilation (depicted as black lines) \cite{PDG20}.}
\label{spectrumXipQ}
\end{figure*}

\begin{figure*}[h]
\centering
\begin{minipage}{\linewidth}
\includegraphics[width=0.47\linewidth]{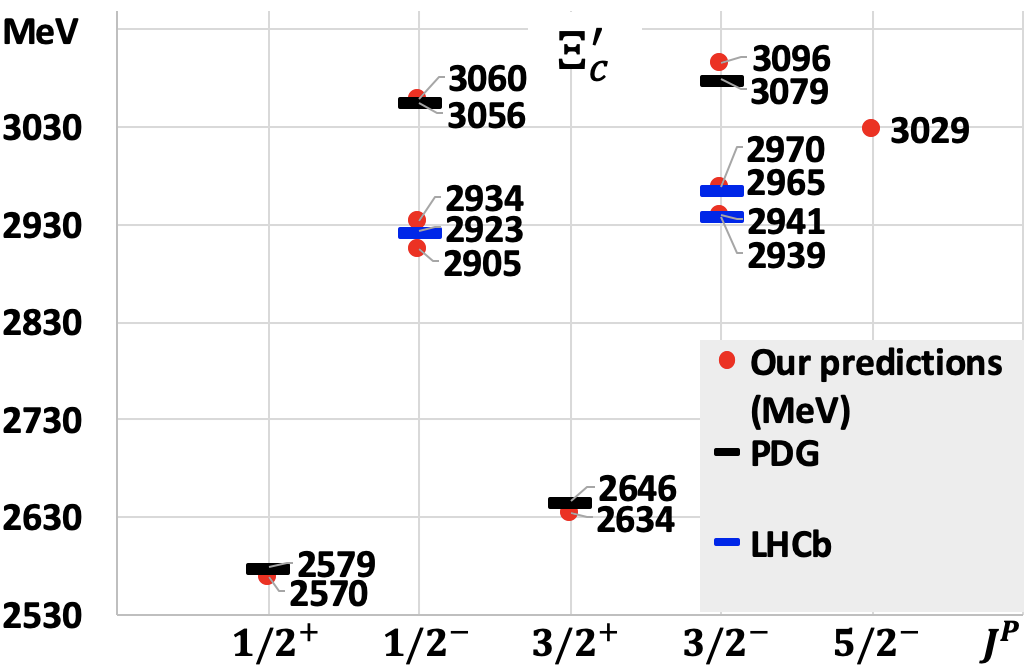}
\hfill
\includegraphics[width=0.47\linewidth]{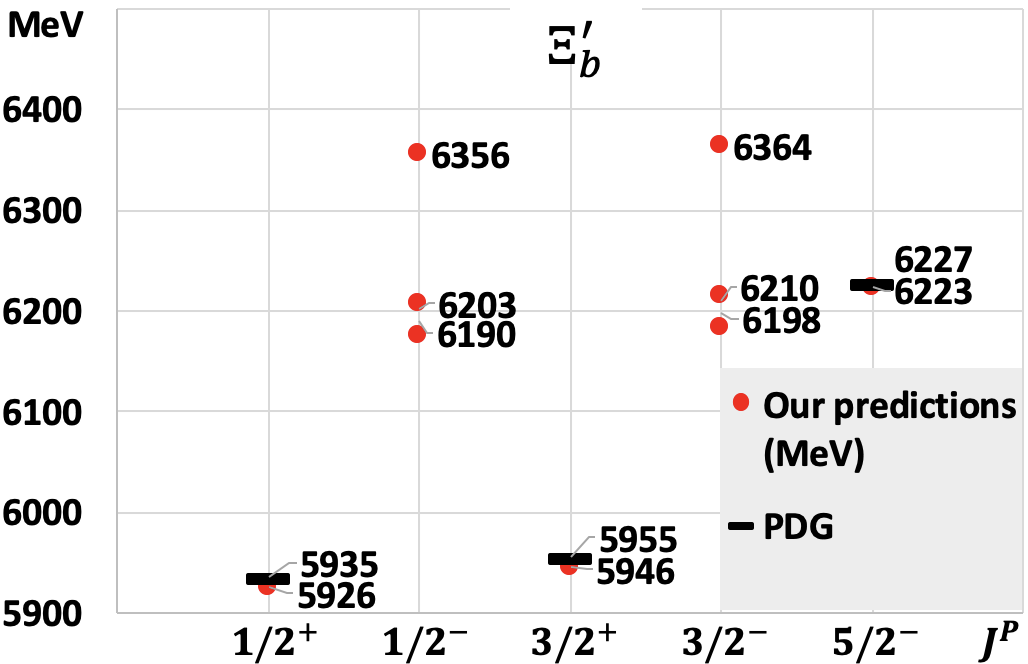}
\end{minipage}
\caption{The mass spectra of $\Xi'_c$ (left) and $\Xi'_b$ (right)  along with tentative quantum number assignments, are shown. The theoretical predictions (represented by red dots) are compared with experimental results obtained by the LHCb collaboration \cite{PhysRevLett.124.222001} \cite{PhysRevLett.128.162001} (depicted as blue lines), as well as with the Particle Data Group compilation (depicted as black lines) \cite{PDG20}.}
\label{spectrumXiQ}
\end{figure*}

The values of the parameters in the mass formula of Eq.~(\ref{hosc}) were determined by studying some well established singly-heavy baryons \cite{Santopinto2019}\cite{Ortiz-Pacheco:2020hmj}.  
In Figs.~\ref{spectrumXipQ} and \ref{spectrumXiQ} we present the mass spectra of the $\Xi_{c/b}$ and the $\Xi'_{c/b}$ baryons, respectively. We compare our results (in red points) with the available experimental data (in black and blue lines) \cite{PDG20}.  We find a good agreement between the experiment and our calculated theoretical masses for the majority of the $\Xi_{c/b}$ and $\Xi'_{c/b}$ baryons. The explicit assignments between the data and our states can be seen in Tables~\ref{dwc} and \ref{dwb} \cite{PhysRevD.105.074029}. 

In the same tables 
we show the results for the strong decays in two models: the elementary emission model (EEM) and the
$^3P_0$ model. The values correspond to the total strong decay widths. In general we find a reasonable agreement between our calculation of the strong decay widths with the experimental data.  

\begin{table}[t]
\centering
\caption{Comparison between the strong total decay widths for charmed baryons $\Xi_c$ and $\Xi_c^{'}$  calculated within EEM and $^3P_0$ model with  the chiral quark model \cite{Wang2017} and the available experimental data. All values are expressed in MeV.}
\label{dwc}
\vspace{10pt}
\begin{tabular}{crrccc}
\hline\hline
\noalign{\smallskip}
& \multicolumn{2}{c}{This work} & $\chi$QM & Exp & Baryon \\
State & EEM & $^3P_0$ & \cite{Wang2017}& \cite{PDG20} & \\
\noalign{\smallskip}
\hline 
\noalign{\smallskip}
$^4\Xi'_{c}$                    &  1.23 &  0.02 &       & $2.14 \pm 0.19$ 
& $\Xi_c(2645)^+$ \\            &  1.23 &  0.02 &       & $2.35 \pm 0.22$ 
& $\Xi_c(2645)^0$ \\
$^2\lambda(\Xi'_{c})_{{1/2}^-}$ &  1.75 &  0.79 & 21.67 && \\
$^4\lambda(\Xi'_{c})_{{1/2}^-}$ &  4.83 &  0.53 & 37.05 & $ 7.1 \pm 2.0$ 
& $\Xi_c(2923)^0$ \\
$^2\lambda(\Xi'_{c})_{{3/2}^-}$ & 11.53 &  3.08 & 20.89 & $14.8 \pm 9.1$ 
& $\Xi_c(2930)^+$ \\            & 11.53 &  3.08 &       & $10.2 \pm 1.4$ 
& $\Xi_c(2939)^0$ \\
$^4\lambda(\Xi'_{c})_{{3/2}^-}$ &  3.88 &  2.04 & 12.33 & $14.1 \pm 1.6$ 
& $\Xi_c(2965)^0$ \\
$^4\lambda(\Xi'_{c})_{{5/2}^-}$ & 29.27 &  5.43 & 20.20 &&  \\
$^2\rho(\Xi'_{c})_{{1/2}^-}$    & 10.73 &  6.26 &       & $ 7.8 \pm 1.9$ 
& $\Xi_c(3055)^+$ \\
$^2\rho(\Xi'_{c})_{{3/2}^-}$    & 25.39 &  3.70 &       & $ 3.6 \pm 1.1$ 
& $\Xi_c(3080)^+$ \\            & 25.39 &  3.70 &       & $ 5.6 \pm 2.2$ 
& $\Xi_c(3080)^0$ \\
\noalign{\smallskip}
\hline
\noalign{\smallskip}
$^2\lambda(\Xi_{c})_{{1/2}^-}$ &  0.01 & 0.41 & 3.61 & $8.9  \pm 1.0$ 
& $\Xi_c(2790)^+$ \\           &  0.01 & 0.41 && $10.0 \pm 1.1$  & $\Xi_c(2790)^0$ \\
$^2\lambda(\Xi_{c})_{{3/2}^-}$ &  0.28 & 0.54 & 2.11 & $2.43 \pm 0.26$ 
& $\Xi_c(2815)^+$ \\           &  0.28 & 0.54 && $2.54 \pm 0.25$ & $\Xi_c(2815)^0$ \\
$^2\rho(\Xi_{c})_{{1/2}^-}$    &  4.57 & 0.70 &&& \\
$^4\rho(\Xi_{c})_{{1/2}^-}$    & 10.20 & 0.45 &&& \\
$^2\rho(\Xi_{c})_{{3/2}^-}$    & 19.45 & 3.76 &&& \\
$^4\rho(\Xi_{c})_{{3/2}^-}$    &  7.60 & 2.51 &&& \\
$^4\rho(\Xi_{c})_{{5/2}^-}$    & 44.91 & 4.55 &&& \\
\noalign{\smallskip}
\hline\hline 
\end{tabular}
\end{table}

\begin{table}[h]
\centering
\caption{As in Tab. \ref{dwc} but for $\Xi_b$ and $\Xi_b^{'}$.}
\label{dwb}
\vspace{10pt}
\begin{tabular}{crrccc}
\hline\hline
\noalign{\smallskip}
& \multicolumn{2}{c}{This work} & $\chi$QM & Exp & Baryon \\
State & EEM & $^3P_0$ & \cite{Wang2017} & \cite{PDG20} \\
\noalign{\smallskip}
\hline 
\noalign{\smallskip}
$^2\Xi'_{b}$                    & $-$   & $-$  &  0.08 & $<0.08$ \\
$^4\Xi'_{b}$                    &  0.48 & 0.02 &  0.98 & $0.90 \pm 0.18$ 
& $\Xi_b(5945)^0$ \\            &  0.48 & 0.02 &       & $1.65 \pm 0.33$ 
& $\Xi_b(5955)^-$ \\
$^2\lambda(\Xi'_{b})_{{1/2}^-}$ &  1.78 & 0.86 & 27.05 && \\
$^4\lambda(\Xi'_{b})_{{1/2}^-}$ &  3.99 & 0.65 & 32.24 && \\
$^2\lambda(\Xi'_{b})_{{3/2}^-}$ &  8.30 & 2.92 & 24.15 && \\
$^4\lambda(\Xi'_{b})_{{3/2}^-}$ &  2.33 & 1.83 & 15.83 && \\
$^4\lambda(\Xi'_{b})_{{5/2}^-}$ & 13.80 & 3.36 & 24.39 & $18.6^{+5.2}_{-4.3}$ 
& $\Xi_b(6227)^0$ \\            & 13.80 & 3.36 &       & $19.9 \pm 2.6$ 
& $\Xi_b(6227)^-$ \\
$^2\rho(\Xi'_{b})_{{1/2}^-}$    & 10.93 & 5.88 &&& \\
$^2\rho(\Xi'_{b})_{{3/2}^-}$    & 13.83 & 3.08 &&& \\
\noalign{\smallskip}
\hline
\noalign{\smallskip}
$^2\lambda(\Xi_{b})_{{1/2}^-}$ & $-$   & $-$    & 2.88 && \\
$^2\lambda(\Xi_{b})_{{3/2}^-}$ & $-$   & $-$    & 2.95 && \\
$^2\rho(\Xi_{b})_{{1/2}^-}$    &  1.94 & 0.55 &&& \\
$^4\rho(\Xi_{b})_{{1/2}^-}$    &  4.21 & 0.36 &&& \\
$^2\rho(\Xi_{b})_{{3/2}^-}$    &  8.44 & 1.90 &&& \\
$^4\rho(\Xi_{b})_{{3/2}^-}$    &  2.47 & 1.90 &&& \\
$^4\rho(\Xi_{b})_{{5/2}^-}$    & 13.45 & 2.16 &&& \\
\noalign{\smallskip}
\hline\hline 
\end{tabular}
\end{table}

Finally, we present our results of the electromagnetic decay widths for the $S$- and $P$-wave single charm baryons. In Table~\ref{emxic} we present a comparison with the experimental data on radiative decay widths of the $\Xi_c(2790)$ and $\Xi_c(2915)$ baryons \cite{PhysRevD.102.071103}.  The Belle collaboration found that the decay width for the neutral baryons is much larger than that for the charged baryons (albeit with large error bars), in agreement with the present calculations as well as with the results of the chiral quark model \cite{Wang2017}, thus confirming the assignment of these statas as a $\lambda$-mode excitation. In Tables~\ref{radexcc} and \ref{radexcb} we show the electromagnetic decay widths for $S$- and $P$-wave single-heavy cascade baryons. The results are in qualitative agreement with those obtained in the chiral quark model. 

\begin{table}[t]
\centering
\caption{Radiative decay widths of $\Xi_c(2790)$ and $\Xi_c(2815)$ baryons in keV.}
\vspace{10pt}
\label{emxic}
\begin{tabular}{crcc}
\hline\hline
\noalign{\smallskip}
Decay & Our& $\chi$QM & Exp \\
& work & \cite{Wang2017} & \cite{PhysRevD.102.071103} \\
\noalign{\smallskip}
\hline
\noalign{\smallskip}
$\Xi_{c}(2790)^+ \rightarrow {}^2\Xi_{c}^+ + \gamma$ 
& $5.4$   & $4.6$   & $<350$ \\
$\Xi_{c}(2790)^0 \rightarrow {}^2\Xi_{c}^0 + \gamma$ 
& $239.3$ & $263.0$ & $800 \pm 320$ \\
$\Xi_{c}(2815)^+ \rightarrow {}^2\Xi_{c}^+ + \gamma$ 
& $2.4$   & $2.8$   & $<80$ \\
$\Xi_{c}(2815)^0 \rightarrow {}^2\Xi_{c}^0 + \gamma$ 
& $344.6$ & $292.0$ & $320 \pm 45^{+45}_{-80}$ \\
\noalign{\smallskip}
\hline\hline
\end{tabular}
\end{table}

\begin{table}[t]
\centering
\caption{Radiative decay widths of $S$- and $P$-wave $\Xi'_c$ and $\Xi_c$ baryons in keV.}
\vspace{10pt}
\label{radexcc}
\begin{tabular}{crrrrrrl}
\hline\hline
\noalign{\smallskip}
& \multicolumn{2}{c}{$^2\Xi'_{c} +\gamma$} 
& \multicolumn{2}{c}{$^4\Xi'_{c} +\gamma$} 
& \multicolumn{2}{c}{$^2\Xi_{c} +\gamma$} & \\ 
& $+$ & $0$ & $+$ & $0$ & $+$ & $0$ & \\ 
\noalign{\smallskip}
\hline
\noalign{\smallskip}
$^2\Xi'_{c}$ & & & & &  15.1 & 0.3 & \\
             & & & & &  42.3 & 0.0 & \cite{Wang2017} \\
$^4\Xi'_{c}$ & 0.0 & 0.9 & & &  59.8 & 1.3 & \\
             & 0.0 & 3.0 & & & 139 & 0.0 & \cite{Wang2017} \\
$^2\lambda(\Xi'_{c})_{1/2^-}$ &  0.4 & 183.5 &  0.4 &   0.2 &  42.9 &  0.9 & \\
                              &  0.0 & 472.0 &  1.6 &   1.0 &  46.4 &  0.0 & \cite{Wang2017} \\
$^2\lambda(\Xi'_{c})_{3/2^-}$ & 17.0 & 401.7 &  0.7 &   0.3 &  57.5 &  1.2 & \\
                              & 12.1 & 302.0 &  1.6 &   1.0 &  46.1 &  0.0 & \cite{Wang2017} \\
$^4\lambda(\Xi'_{c})_{1/2^-}$ &  1.5 &   0.7 &  0.3 &  20.3 &  27.2 &  0.6 & \\
                              &  0.3 &   0.2 &  0.2 & 125.0 &  14.5 &  0.0 & \cite{Wang2017} \\
$^4\lambda(\Xi'_{c})_{3/2^-}$ &  6.5 &   2.8 &  0.5 & 122.4 &  99.4 &  2.1 & \\
                              &  2.1 &   1.2 &  1.6 & 187.0 &  54.6 &  0.0 & \cite{Wang2017} \\
$^4\lambda(\Xi'_{c})_{5/2^-}$ &  7.3 &   2.9 & 12.0 & 293.2 &  92.9 &  2.0 & \\
                              &  1.6 &   0.9 &  2.3 & 192.0 &  32.0 &  0.0 & \cite{Wang2017} \\
$^2\rho(\Xi'_{c})_{1/2^-}$    & 16.3 &  26.6 &  4.7 &   7.7 & 731.2 & 15.5 & \\
$^2\rho(\Xi'_{c})_{3/2^-}$    & 20.9 &  34.3 &  6.4 &  10.5 & 783.4 & 16.6 & \\
\noalign{\smallskip}
\hline
\noalign{\smallskip}
$^2\lambda(\Xi_{c})_{1/2^-}$ &   2.3 &  0.0 &   0.2 & 0.0 &  5.4 & 239.3 & \\
                             &   1.4 &  0.0 &   0.4 & 0.0 &  4.6 & 263.0 & \cite{Wang2017} \\
$^2\lambda(\Xi_{c})_{3/2^-}$ &   4.6 &  0.1 &   0.6 & 0.0 &  2.4 & 344.6 & \\
                             &   2.3 &  0.0 &   1.0 & 0.0 &  2.8 & 292.0 & \cite{Wang2017} \\
$^2\rho(\Xi_{c})_{1/2^-}$    & 157.2 &  3.3 &   1.8 & 0.0 & 16.0 &  26.2 & \\
                             & 128.0 &  0.0 &   0.2 & 0.0 &  1.4 &   5.6 & \cite{Wang2017} \\
$^2\rho(\Xi_{c})_{3/2^-}$    & 585.1 & 12.4 &   2.8 & 0.1 & 20.6 &  33.7 & \\
                             & 110.0 &  0.0 &   0.5 & 0.0 &  1.9 &   7.5 & \cite{Wang2017} \\
$^4\rho(\Xi_{c})_{1/2^-}$    &   5.4 &  0.1 &  12.5 & 0.3 &  9.8 &  16.1 & \\
                             &   0.4 &  0.0 &  43.4 & 0.0 &  0.7 &   3.0 & \cite{Wang2017} \\
$^4\rho(\Xi_{c})_{3/2^-}$    &  21.2 &  0.4 & 122.4 & 2.6 & 34.6 &  56.6 & \\
                             &   1.8 &  0.0 &  58.1 & 0.0 &  2.8 &  11.2 & \cite{Wang2017} \\
$^4\rho(\Xi_{c})_{5/2^-}$    &  21.7 &  0.5 & 445.5 & 9.5 & 30.8 &  50.5 & \\
\noalign{\smallskip}
\hline\hline
\end{tabular}
\end{table}

\begin{table}[t]
\centering
\caption{Radiative decay widths of $S$- and $P$-wave $\Xi'_b$ and $\Xi_b$ 
baryons in keV.}
\vspace{10pt}
\label{radexcb}
\begin{tabular}{crrrrrrl}
\hline\hline
\noalign{\smallskip}
& \multicolumn{2}{c}{$^2\Xi'_{b} +\gamma$} 
& \multicolumn{2}{c}{$^4\Xi'_{b} +\gamma$} 
& \multicolumn{2}{c}{$^2\Xi_{b} +\gamma$} & \\ 
& $0$ & $-$ & $0$ & $-$ & $0$ & $-$ & \\ 
\noalign{\smallskip}
\hline
\noalign{\smallskip}
$^2\Xi'_{b}$ & & & & &  33.2 & 0.7 & \\
             & & & & &  84.6 & 0.0 & \cite{Wang2017} \\
$^4\Xi'_{b}$ & 0.0 & 0.0 & & &  49.5 & 1.0 & \\
             & 5.2 & 15.0 & & & 104 & 0.0 & \cite{Wang2017} \\
$^2\lambda(\Xi'_{b})_{1/2^-}$ &  48.2 &  46.4 &  0.3 &   0.5 &  53.7 &  1.1 & \\
                              &  76.3 & 190.0 &  0.9 &   3.5 &  72.2 &  0.0 & \cite{Wang2017} \\
$^2\lambda(\Xi'_{b})_{3/2^-}$ & 101.2 & 116.0 &  0.3 &   0.5 &  57.9 &  1.2 & \\
                              &  43.9 &  92.3 &  0.9 &   3.6 &  72.8 &  0.0 & \cite{Wang2017} \\
$^4\lambda(\Xi'_{b})_{1/2^-}$ &   0.5 &   0.8 &  6.5 &   5.3 &  30.3 &  0.6 & \\
                              &   0.3 &   1.5 & 69.5 & 164.0 &  34.0 &  0.0 & \cite{Wang2017} \\
$^4\lambda(\Xi'_{b})_{3/2^-}$ &   1.4 &   2.5 & 35.6 &  35.4 &  90.2 &  1.9 & \\
                              &   0.7 &   2.0 & 47.5 & 104.0 &  94.0 &  0.0 & \cite{Wang2017} \\
$^4\lambda(\Xi'_{b})_{5/2^-}$ &   1.1 &   1.9 & 64.1 &  76.1 &  64.8 &  1.4 & \\
                              &   0.4 &   1.9 & 41.5 &  88.2 &  47.7 &  0.0 & \cite{Wang2017} \\
$^2\rho(\Xi'_{b})_{1/2^-}$    &  12.3 &  20.2 &  5.2 &   8.6 & 698.5 & 14.8 & \\
$^2\rho(\Xi'_{b})_{3/2^-}$    &  13.2 &  21.6 &  5.6 &   9.2 & 705.7 & 15.0 & \\
\noalign{\smallskip}
\hline
\noalign{\smallskip}
$^2\lambda(\Xi_{b})_{1/2^-}$ &   0.2 &  0.0 &   0.0 & 0.0 & 72.8 &  80.0 & \\
                             &   1.3 &  0.0 &   2.0 & 0.0 & 63.6 & 135.0 & \cite{Wang2017} \\
$^2\lambda(\Xi_{b})_{3/2^-}$ &   0.3 &  0.0 &   0.1 & 0.0 & 79.3 &  85.6 & \\
                             &   1.7 &  0.0 &   2.6 & 0.0 & 68.3 & 147.0 & \cite{Wang2017} \\
$^2\rho(\Xi_{b})_{1/2^-}$    & 120.9 &  2.6 &   1.2 & 0.0 & 11.9 &  19.5 & \\
                             &  94.3 &  0.0 &   0.6 & 0.0 &  1.9 &   7.2 & \cite{Wang2017} \\
$^2\rho(\Xi_{b})_{3/2^-}$    & 296.9 &  6.3 &   1.3 & 0.0 & 12.9 &  21.1 & \\
                             &  69.4 &  0.0 &   0.8 & 0.0 &  2.1 &   8.1 & \cite{Wang2017} \\
$^4\rho(\Xi_{b})_{1/2^-}$    &   1.9 &  0.0 &  13.7 & 0.3 &  6.7 &  10.9 & \\
                             &   0.2 &  0.0 &  80.0 & 0.0 &  0.9 &   3.6 & \cite{Wang2017} \\
$^4\rho(\Xi_{b})_{3/2^-}$    &   5.9 &  0.1 &  91.5 & 1.9 & 19.9 &  32.5 & \\
                             &   0.8 &  0.0 &  78.0 & 0.0 &  2.9 &  11.4 & \cite{Wang2017} \\
$^4\rho(\Xi_{b})_{5/2^-}$    &   4.6 &  0.1 & 193.9 & 4.1 & 14.1 &  23.1 & \\
\noalign{\smallskip}
\hline\hline
\end{tabular}
\end{table}

\section{Summary and conclusions}

We constructed the low-lying baryon states $\Xi_{c/b}$ and  $\Xi'_{c/b}$ in the quark model for $S$- and $P$-wave and make an analysis related with their quantum numbers and the available experimental data. The assignment of quantum numbers to the experimentally observed single-charm and single-bottom baryons was based on a combination of energy systematics, strong and electromagnetic decay widths. Overall, a reasonable agreement was found with the available experimental data.  

\ack
It is a pleasure to thank Elena Santopinto, Alessandro Giachino and Hugo Garc{\'{\i}}a-Tecocoatzi for interesting discussions. This work was supported by grant No.~IG101423 from PAPIIT, DGAPA-UNAM, Mexico, and by grant No.~251817 from CONACyT, Mexico. EOP acknowledges the projects (P1-0035) and (J1-3034) with financial support from the Slovenian Research Agency.

\section*{References}

%\bibliographystyle{apsrev4-1}
% \bibliography{biblio.bib}
%\input{Heavybaryon.bbl}

\bibliography{Xi_baryons}

\providecommand{\newblock}{}
\begin{thebibliography}{1}
\expandafter\ifx\csname url\endcsname\relax
  \def\url#1{{\tt #1}}\fi
\expandafter\ifx\csname urlprefix\endcsname\relax\def\urlprefix{URL }\fi
\providecommand{\eprint}[2][]{\url{#2}}

\bibitem{PDG20}
Zyla P {\em et~al.\/} (Particle Data Group) 2020 {\em PTEP\/} {\bf 2020} 083C01

\bibitem{Isgurandkarl(1978)}
Isgur N and Karl G 1978 {\em Phys. Rev. D\/} {\bf 18}(11) 4187--4205

\bibitem{Santopinto2019}
Santopinto E, Giachino A, Ferretti J, Garc{\'\i}a-Tecocoatzi H, Bedolla M~A,
  Bijker R and Ortiz-Pacheco E 2019 {\em The European Physical Journal C\/}
  {\bf 79} 1012

\bibitem{Ortiz-Pacheco:2020hmj}
Ortiz-Pacheco E, Bijker R, Giachino A and Santopinto E 2020 {\em J. Phys. Conf.
  Ser.\/} {\bf 1610} 012011 (\textit{Preprint} \eprint{2004.09409})

\bibitem{PhysRevD.105.074029}
Bijker R, Garc\'{\i}a-Tecocoatzi H, Giachino A, Ortiz-Pacheco E and Santopinto
  E 2022 {\em Phys. Rev. D\/} {\bf 105}(7) 074029

\bibitem{PhysRevLett.124.222001}
Aaij R {\em et~al.\/} (LHCb Collaboration) 2020 {\em Phys. Rev. Lett.\/} {\bf
  124}(22) 222001

\bibitem{PhysRevLett.128.162001}
Aaij R {\em et~al.\/} (LHCb Collaboration) 2022 {\em Phys. Rev. Lett.\/} {\bf
  128}(16) 162001

\bibitem{Wang2017}
Wang K~L, Yao Y~X, Zhong X~H and Zhao Q 2017 {\em Phys. Rev. D\/} {\bf 96}(11)
  116016

\bibitem{PhysRevD.102.071103}
Yelton J {\em et~al.\/} (The Belle Collaboration) 2020 {\em Phys. Rev. D\/}
  {\bf 102}(7) 071103

\end{thebibliography}

\end{document}